# Massless Dirac Fermions Trapping in a Quasi-one-dimensional *npn* Junction of a Continuous Graphene Monolayer

Ke-Ke Bai[1], Jia-Bin Qiao[1], Hua Jiang[2,*], Haiwen Liu[1,*], Lin He[1,*]

[1]Center for Advanced Quantum Studies, Department of Physics, Beijing Normal University, Beijing, 100875, People's Republic of China

[2]College of Physics, Optoelectronics and Energy, Soochow University, Suzhou, 215006, People's Republic of China

[§]These authors contributed equally to this work.

*Correspondence and requests for materials should be addressed to H.J. (e-mail: jianghuaphy@suda.edu.cn), H.L. (haiwen.liu@bnu.edu.cn), and L.H. (e-mail: helin@bnu.edu.cn).

**Massless Dirac fermions in graphene provide unprecedented opportunities to realize the Klein paradox, which is one of the most exotic and striking properties of relativistic particles. In the seminal theoretical work [Katsnelson et al., *Nat. Phys.* 2 620 (2006)], it was predicted that the massless Dirac fermions can pass through one-dimensional (1D) potential barriers unimpededly at normal incidence. Such a result seems to preclude confinement of the massless Dirac fermions in graphene by using 1D potential barriers. Here, we demonstrate, both experimentally and theoretically, that massless Dirac fermions can be trapped in quasi-1D *npn* junction of a continuous graphene monolayer. Because of highly anisotropic transmission of the massless Dirac fermions at *n-p* junction boundaries (the so-called Klein tunneling in graphene), charge carries incident at large oblique angles will be reflected from one edge of the junction with high probability and continue to bounce from the opposite edge. Consequently, these electrons are trapped for a finite time to form quasi-bound states in the quasi-1D *npn* junction. The quasi-bound states seen as pronounced resonances are probed and the quantum interference patterns arising from these states are directly visualized in our scanning tunneling microscope measurements.**

In the seminal theoretical work of Katsnelson, et al. [1], the authors demonstrated the perfect transmission for massless Dirac fermions incident in the normal direction of a one-dimensional (1D) potential barrier in graphene monolayer. This is treated as a solid-state analogue [1-8] to the realization of the so-called Klein paradox, a relativistic tunneling effect that was first discussed by Klein for 1D potentials [9]. Such a phenomenon leads to completely confinement of the massless Dirac fermions in graphene by 1D potential barrier impossible. Inspired by the classical whispering-gallery mode physics in optics and acoustics, theorists proposed the confinement of the massless Dirac fermions in graphene using circular graphene *p-n* junctions [10-14]. Until very recently, scientists developed effective methods to generate high-quality circular *p-n* junctions in graphene and were able to study this subject experimentally [15-17]. Experiments, using scanning tunneling microscope (STM), show that the massless Dirac fermions can temporarily be trapped inside the circular *p-n* junctions and wavefunctions of the charge carriers are confined as quantified modes by continuous reflection near the interface of closed circle potential barriers [15-17]. These results directly demonstrated the unusual anisotropic transmission of the massless Dirac fermions at the *p-n* junction boundaries, *i.e.*, the so-called Klein tunneling in graphene [1], at the atomic scale. However, an atomic-scale verification of the Klein tunneling using nanoscale 1D potential barriers (the first proposed structure in the initial theoretical work [1,9]) is still missing up to now, not even to mention direct mapping of the confined massless Dirac fermions in the 1D junctions of graphene.

In this Letter, we address the above issue and demonstrate that massless Dirac fermions can be trapped in quasi-1D *npn* junctions of graphene created by substrate engineering. Different from the cylindrical symmetry of circular *pn* junctions, where the confined electron states are described by radial quantum number *n* and azimuthal quantum number *m* [15-17], the confined electron states in the 1D *npn* junction are highly anisotropy. Via STM measurements, we directly probe the quasi-bound states and image the wavefunctions of the confined massless Dirac fermions within the 1D *npn* junction. All the experimental features are reproduced quite well in our theoretical calculations. Our experiments not only shed light on the Klein tunneling, but also demonstrate a facile and new method for generating 1D electronic junctions in graphene.

In our experiment, graphene monolayer was grown on a Cu foil by chemical vapor deposition (CVD) method (see Supplemental Material [18] for details) [19]. The Cu foil was annealed in 1000 °C to form large scale single-crystal Cu surface, and then graphene monolayer was grown on it [20,21]. Fig. 1(a) shows a representative large-scale atomic force microscope (AFM) image of graphene monolayer across Cu steps. Usually, the edges of the Cu terraces are quite straight and long (they are usually longer than 1 μm). The X-ray diffraction measurement of the Cu foil, as shown in Fig. S1, reveals that the exposed Cu terrace are usually {100}, {110}, {111} and {311} surfaces. Such a result is further verified in the zoom-in STM measurements. We observe different topographic moiré patterns due to the misorientations between the graphene and the exposed Cu lattices, as shown in Fig. S2 [18]. Our quasi-1D *npn* junctions are generated along the edges between two adjacent Cu terraces because of that there are different potential energies for graphene along the edges and on the terraces. Fig. 1(b) shows an enlarged STM image of a typical edge between two large Cu terraces. The height between the two terraces is about 2.8 nm and a series of mini facets can be seen between the two terraces, as schematically shown in Fig. 1(c) (see Supplemental Material Fig.S3 [18] for details). For clarity, we divide the studied structure into three regions and label them as I, II, and III from left to right. The distance between graphene and Cu surface is expected to be slightly different for graphene along the edges (the region II) and on the terraces (the regions I and III). Variation of graphene-Cu separations, i.e., the overlap of graphene and Cu wave functions, can affect the position of the Dirac point in graphene [17,22]. By simply increasing the graphene-Cu separations, the graphene can be changed from *n*-doped to *p*-doped at the atomic scale. Therefore, such a structure with various graphene-metal separations provides an ideal platform to realize the quasi-1D *npn* junctions in graphene.

Fig. 2(a) shows five typical scanning tunneling spectroscopy (STS) spectra recorded at different positions across the edge of Cu terraces shown in Fig. 1(b). The tunneling spectrum gives direct access to the local density of states (LDOS) of the surface at the position of the STM tip. A local minimum of the tunneling conductance, as pointed out by the arrows in Fig. 1(a), is attributed to the Dirac point, $E_D$, of graphene. The local Dirac point recorded on the left terrace (the region I), ~ 2 nm away from the left edge, is about -320 meV, indicating *n*-doping. It shifts abruptly to above 400 meV (*p*-doping) in the region II, indicating effect of the graphene-Cu interaction on the charge transfer between them. The Dirac point shifts back to negative energies, at about -160 meV (*n*-doping), on the right terrace (the region III). Further spatial-resolved STS measurements across the *npn* junction, as shown in Fig. S4 [18], indicates that the potential barriers of the *n-p* and *p-n* junctions are atomically sharp, with the width of only about 0.5 nm. The result in Fig. S4 also confirms that the entire region II is *p*-doping, regardless of small variation of Dirac point in this region. Therefore, we realize a quasi-1D *npn* junction along the edge of the Cu terrace, as schematically shown in Fig. 1(c).

The most striking feature that we observed inside the quasi-1D *npn* junction is the appearance of a series of almost equally spaced resonances in the tunneling conductance (Fig. 2(a)). The average energy spacing of the resonances is about 46 meV (see Supplemental Material Fig.S5 [18]). The spatial variation of the tunneling spectra, as shown in Fig. 2(a) and Fig. S3, precludes any possible artificial effects of the STM tips as the origin of these features. We find that these resonant peaks are much pronounced for energies below the Dirac point of the *p*-doped region. For 1D potential barrier in a graphene monolayer, although theorists predicted perfect transmission for normal incidence of charge carriers, they also showed that quasi-particles incident at large oblique angles will be reflected from the potential barrier with high probability [1,7]. The reflected charge carriers continue to bounce in the quasi-1D *npn* junction many times before escape finally, as schematically shown in Fig. 2(b). These reflected quasi-particles form quasi-bound states in the quasi-1D *npn* junction and give rise to resonances of finite trapping time [15-17]. Therefore, the resonant peaks in the tunneling spectra (Fig. 2(a)) are attributed to the formation of quasi-bound states in the quasi-1D *npn* junction.

To verify the above assumption, we study the electronic properties of the quasi-1D *npn* junction in theory by two different methods: one is scattering calculation based on low energy effective model, the other is based on lattice Green's function. First, we calculate the electronic structures of the quasi-1D *npn* junction by solving the two-dimensional massless Dirac equation in the presence of an asymmetric 1D step-potential, as schematically shown in Fig. 2(b) (see Supplemental Material

[18] for details of calculation). In the calculation, the width of the junction, the energies of the Dirac point and the renormalized Fermi velocities in the three different regions, which are determined experimentally (see Supplemental Material Figs.S3 and S6 [18] for details), are taking into account. The calculated LDOS in the quasi-1D *npn* junctionis shown in Fig. S7 [18]. A series of resonant peaks, with almost evenly energy-spacing ~ 50 meV, emerge in the *p*-doped region of the *npn* junction, which reproduces the main feature of our experimental observations. In the experiment, the potential barriers of the *npn* junction are atomically sharp (Fig. S3) and intervalley mixing is unavoidable around the junction. To include the effect of intervalley scattering on the electronic properties of the junction, we further solve this problem numerically based on the lattice Green's function (see Supplemental Material [18] for details of calculation). Fig. 2(c) shows the theoretical LDOS in the three different regions of the junction. Obviously, our calculation captures well the energies of the resonancesand main features of STS spectra in our experiment (Fig. 2(**a**)). The consistency between the experiment and simulation confirms the formation of quasi-bound states in the quasi-1D *npn* junction of graphene through the Klein tunneling.

To further explore the confined massless Dirac fermions, we carried out measurements of differential conductance maps (STS maps), which reflect the spatial distribution of the LDOS at the recorded energies [16,17]. At a fixed energy, the LDOS at position *r* is determined by the wave-functions according to $LDOS(r) \propto |\psi(r)|^2$ [23-26]. Therefore, the STS maps reflect spatial distribution of the confined Dirac fermions. Fig. 3(a) and 3(c) display two STS maps recorded at different energiesaround the right edge of the *npn* junction (indicated by the blue frame in Fig. 1(b)). Obviously, 1D quantum interference patterns are observed in the *p*-doped region. As the energies increases from -77.2 meV to -61.4 meV, the average wavelength of the interference patterns increases from about 1.6 nm to about 1.9 nm. The calculated STS maps at the corresponding energies, as shown in Fig. 3(b) and Fig. 3(d), reproduce the main features of experimental maps. The main discrepancy between the experimental and the calculated STS maps is that the theoretical interference patterns are more smooth, straight, and equally-spaced. In our experiment, the edge of the Cu terraces (or the edge of the *p*-doped region) is not atomically smooth (Fig. 3(e)), the energy of the Dirac point in the *p*-doped region is not a constant and, importantly, there are atomic defects

(Fig. S8 [18]) and mini facets in the *npn* junction. All these mentioned experimental features, which are not taking into account in our theory, can affect the interference patterns of the quasi-bound states and contribute to the observed discrepancy.

We can also deduce the trapping time $\tau$ of the quasi-bound states via $\tau = \frac{\hbar}{\delta\varepsilon}$, where $\delta\varepsilon$ is the full-width at half-maximum (FWHM) of the resonant peaks in the tunneling spectra and $\hbar$ is Planck's constant divided by $2\pi$ [17,27]. The value of the trapping time is an important parameter to evaluate the effectiveness of the potential barrier in confining the massless Dirac fermions. Fig. 4 shows the trapping time as a function of energies deduced according to the LDOS of the 1D *npn* junction both experimentally and theoretically. Regardless of the larger size of the trapping region in our 1D *npn* junction, the trapping time in our experiment is slightly larger than that of the graphene circular *pn* junction in a recent scanning tunneling microscopy experiment [17]. Such a result indicates that the 1D *npn* junction can trap the massless Dirac fermions as effective as that of the circular *pn* junction. In our theory, we calculated the life time of the quasi-bound states both (a) with considering and (b) without considering the intervalley scattering around the sharp edges of the quasi-1D *npn* junction. It is interesting to note that the intervalley scattering around the edges increases the trapping time of the massless Dirac fermions. Such a result agrees with our basic intuition that the intervalley scattering increases the probability of reflection and, therefore, increases the average times the charge carriers bounced within the 1D junction, i.e., the trapping time of the massless Dirac fermions. According to Fig. 4, the trapping time measured in our experiment is much smaller than that obtained in the theory. This is quite reasonable since that many different effects, which can reduce the trapping time of massless Dirac fermions, are not considered in our calculation. According to our understanding, the most important reason that reduces the trapping time should be the non-zero probability of the massless Dirac fermions tunneling into the surface of the copper substrate. Of course, the thermal and instrumental broadening effect could also reduce the trapping time of the quasi-bound states.

In summary, we demonstrate that the massless Dirac fermions can be locally confined in nanoscale regions of a continuous graphene monolayer by quasi-1D *npn* junctions. It is surprised that the quasi-1D *npn* junction can trap massless Dirac fermions as effective as that of the circular

*pn* junction. The method reported in this work may also provide a facile way to generate nanoscale 1D electronic potential barriers in a continuous graphene monolayer.

**Acknowledgments**

This work was supported by the National Natural Science Foundation of China (Grant Nos. 11674029, 11422430, 11374035, 11374219, 11504008, 11674028), the National Basic Research Program of China (Grants Nos. 2014CB920903, 2013CBA01603, 2014CB920901), the program for New Century Excellent Talents in University of the Ministry of Education of China (Grant No. NCET-13-0054). L.H. also acknowledges support from the National Program for Support of Top-notch Young Professionals.

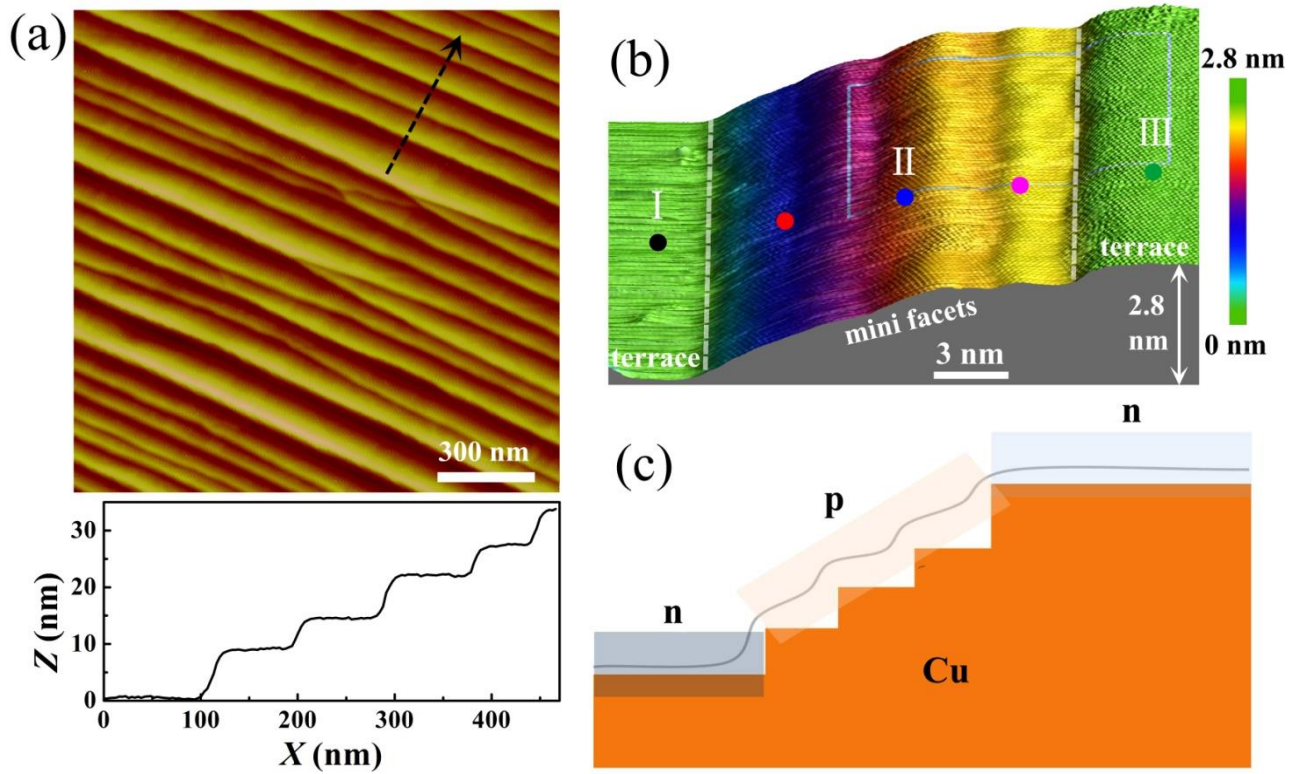

**Figure 1 | (a),** Top panel: A representative large-scale AFM image of graphene monolayer on Cu terraces. Bottom panel: the height profile along the black dashed line in the top panel. (**b**), A typical three-dimensional STM image of graphene monolayer across a Cu step ($V_{sample}$ = -0.614 V and $I$ = 0.154 nA). There are a series of mini facets at the edge of the Cu terrace. We divide the studied structure into three regions (they are separated by two white dashed lines): I and III represent the regions of graphene on Cu terraces, II represents the region of graphene on the edge of the Cu terrace. (**c**), Schematic model of the structure shown in (**b**). The step is composed of a series of mini-facets between the two adjacent Cu terraces. **n** and **p** denote electron-doped and hole-doped regions of the continuous graphene monolayer, respectively.

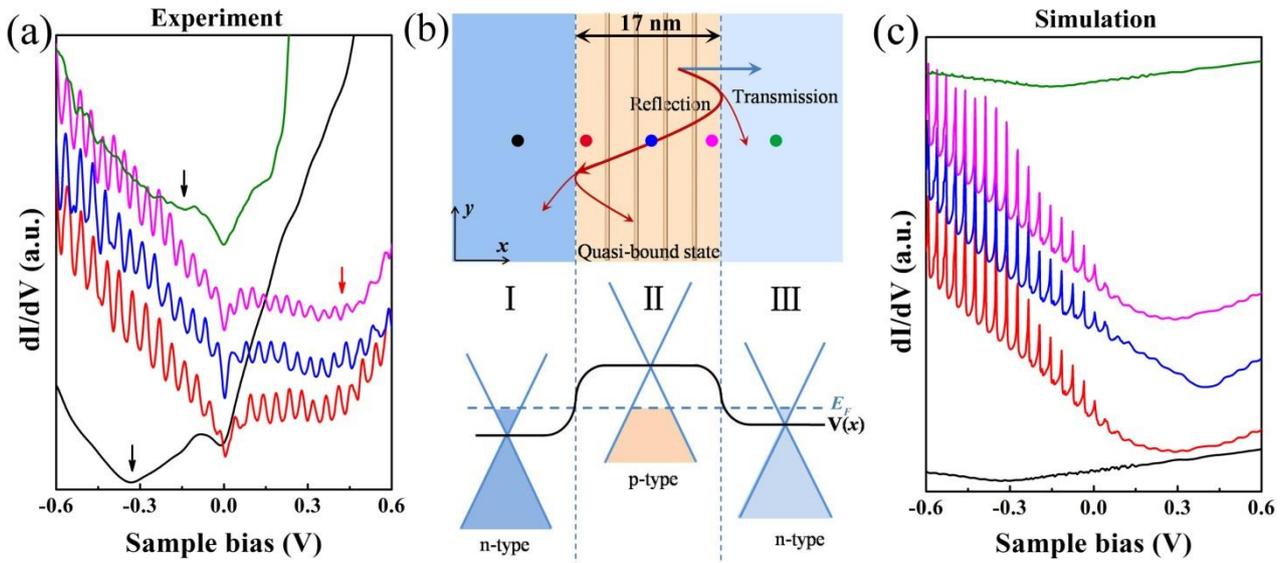

**Figure 2** | (**a**), STS spectra obtained at different positions marked by the dots with different colors in Fig. 1(**b**). The arrows denoted the positions of the Dirac points at different curves in the three different regions. For clarity, the curves are offset in *y*-axis. (**b**), Top: Schematic representation of an 1D n-p-n junction in graphene monolayer. The width of the region II is about 17 nm. The blue arrow indicates the perfect transmission at normal incidence. The dark red curved arrows indicate scattering of charge carriers incident at large oblique angles. The straight lines indicate the standing waves interfered by the incident and reflected waves. Bottom: Schematic diagrams of the spectrum of quasiparticles in graphene monolayer and potential barrier of the quasi-1D *npn* junction. (**c**), The calculated dI/dV curves by lattice Green's function at different positions marked by the solid dots with different colors in the top panel of (**b**).

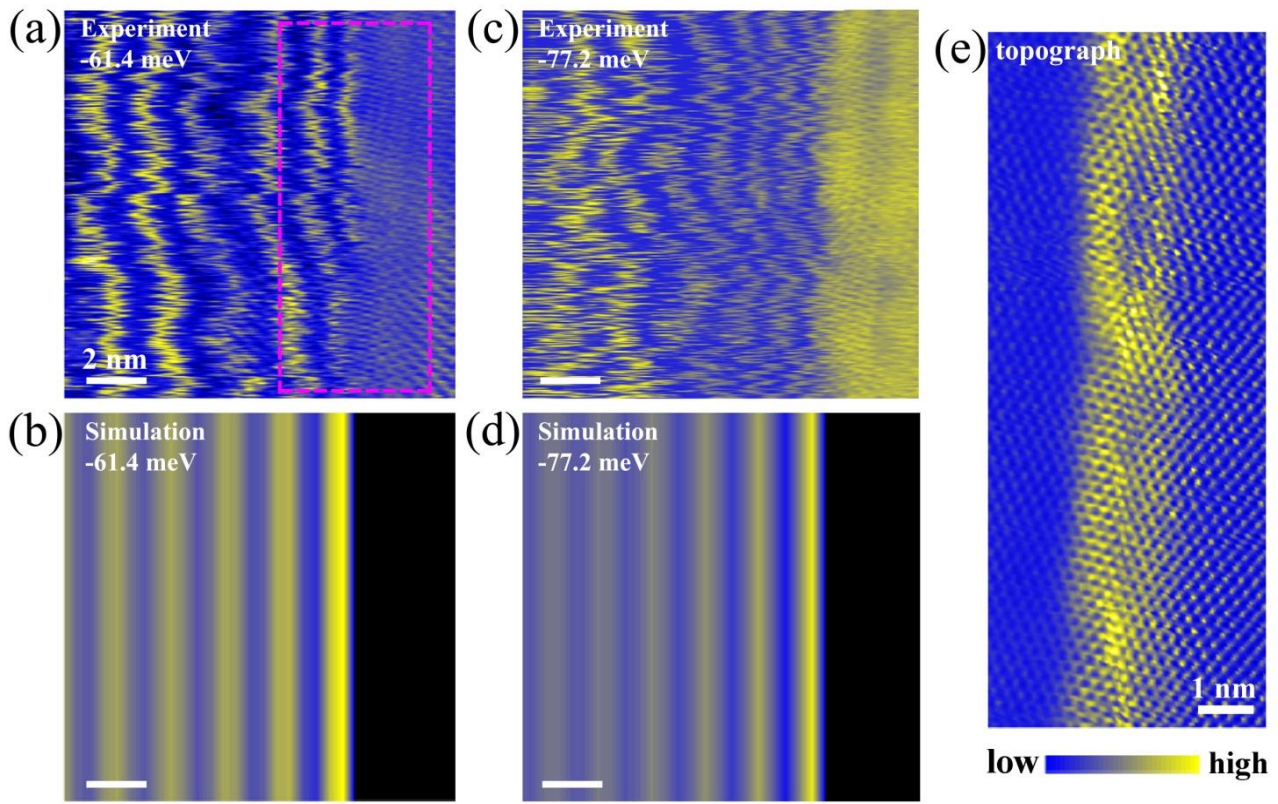

**Figure 3 | (a)** and (**c**), STS maps measured around the right edge of the quasi-1D npn junction in Fig. 1b at energies of -61.4 meV and -77.2 meV, respectively. (**c**) and (**d**), The corresponding STS maps calculated around the right edge of the quasi-1D *npn* junction at energies of -61.4 meV and -77.2 meV, respectively. (**e**), Zoom-in atomic-resolution STM image obtained in the magenta frame in (**a**). The color bar applied to all the LDOS maps and STM image.

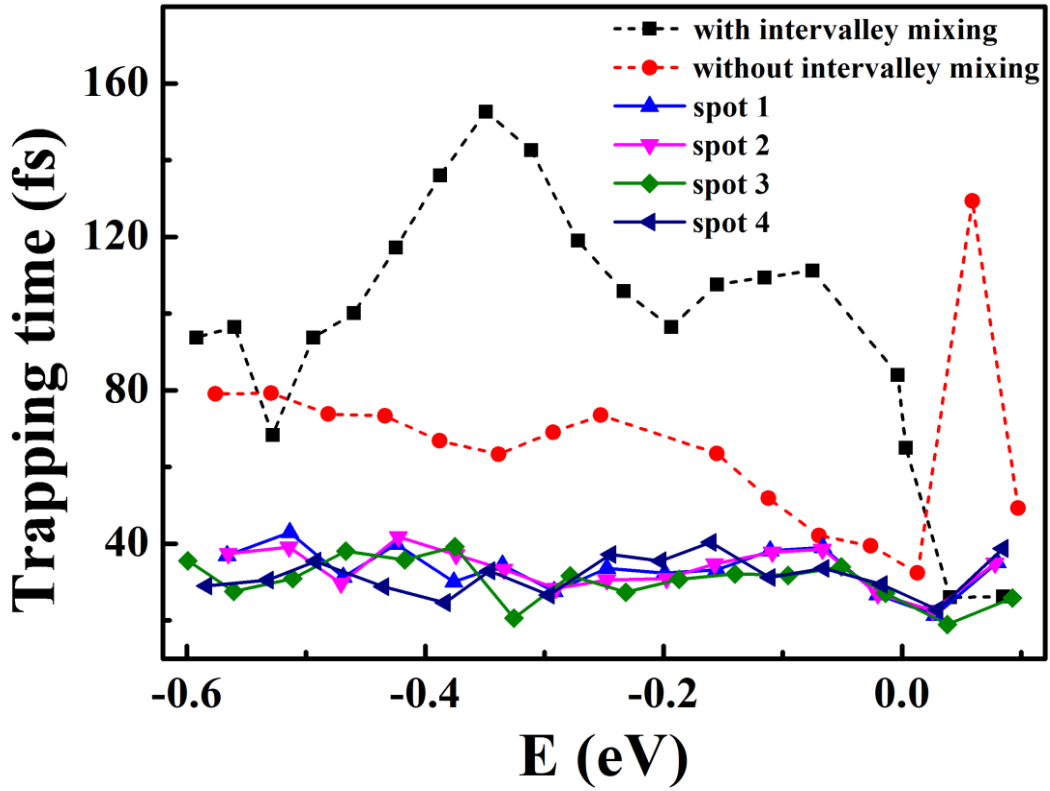

**Figure 4** | The trapping times of the quasi-bound states in the quasi-1D *npn* junction. The trapping times are measured from inverse peak widths of the quasi-bound states both experimentally and theoretically. The trapping times measured in our experiment are smaller than that estimated theoretically. It is also interesting to note that the intervalley scattering of the edges, in general, increases the trapping time of the quasi-bound states.